\documentclass[twocolumn]{article}
\usepackage{graphicx,subfigure,amssymb,amsmath,amsthm,appendix}
\usepackage{hyperref}
\usepackage[margin=0.8in]{geometry}
\usepackage{xcolor}
\usepackage{multirow}
\usepackage{caption}

\DeclareMathOperator*{\maxi}{max}

\begin{document}

\title{Synthesizing Skeletal Motion and Physiological Signals as a Function of a Virtual Human's Actions and Emotions\thanks{A version of this article will appear as \cite{banerjee2021synthesizing}.}}
\author{Bonny Banerjee\thanks{Corresponding author.}~\thanks{Institute for Intelligent Systems, and Department of Electrical \& Computer Engineering, University of Memphis, Memphis, TN 38152, USA. Email: \{bbnerjee, mbaruah\}@memphis.edu.}
\and Masoumeh Heidari Kapourchali\thanks{Department of Electrical \& Computer Engineering, Johns Hopkins University, Baltimore, MD 21218, USA. Email: mheidar1@jhu.edu.}
\and Murchana Baruah\footnotemark[3]
\and Mousumi Deb\thanks{Texas A\&M AgriLife Research \& Extension Center, Temple, TX 76502, USA. Email: mousumi.deb@brc.tamus.edu.}
\and Kenneth Sakauye\thanks{Department of Psychiatry, University of Tennessee Health Science Center, Memphis, TN 38163, USA. Email: ksakauye@uthsc.edu.}
\and Mette Olufsen\thanks{Department of Mathematics, North Carolina State University, Raleigh, NC 27695, USA. Email: msolufse@ncsu.edu.}}

\date{}

\maketitle

\begin{abstract} 
Round-the-clock monitoring of human behavior and emotions is required in many healthcare applications which is very expensive but can be automated using machine learning (ML) and sensor technologies. Unfortunately, the lack of infrastructure for collection and sharing of such data is a bottleneck for ML research applied to healthcare. Our goal is to circumvent this bottleneck by simulating a human body in virtual environment. This will allow generation of potentially infinite amounts of shareable data from an individual as a function of his actions, interactions and emotions in a care facility or at home, with no risk of confidentiality breach or privacy invasion. In this paper, we develop for the first time a system consisting of computational models for synchronously synthesizing skeletal motion, electrocardiogram, blood pressure, respiration, and skin conductance signals as a function of an open-ended set of actions and emotions. Our experimental evaluations, involving user studies, benchmark datasets and comparison to findings in the literature, show that our models can generate skeletal motion and physiological signals with high fidelity. The proposed framework is modular and allows the flexibility to experiment with different models. In addition to facilitating ML research for round-the-clock monitoring at a reduced cost, the proposed framework will allow reusability of code and data, and may be used as a training tool for ML practitioners and healthcare professionals.
\end{abstract}


\section{Introduction} 

\textbf{Motivation:} Individuals with certain mental health conditions, such as Alzheimer’s, schizophrenia, substance abuse, brain injury, trauma, and depression, manifest abnormal or agitated behaviors from time to time. They may require long-term round-the-clock monitoring which is very expensive and strenuous, but can be automated using sensors and machine learning (ML) algorithms. However, collecting real-world data from such patients to train ML algorithms poses five challenges: (1) risk of confidentiality breach and privacy invasion (the patient and his environment has to be monitored simultaneously as the causes behind his behavior may be hidden in his environment), (2) low-level data collection and processing issues (e.g., speech/speaker recognition requires solving the blind signal separation and cocktail party problems), (3) limited resolution and coverage of sensors (e.g., cameras may not be installed in private spaces such as bedrooms and bathrooms), (4) discomfort and lack of motivation to wear sensors round-the-clock, and (5) considerable time and expertise required for data annotation. Unsurprisingly, round-the-clock monitoring data from mentally-ill individuals is scarce. This is a significant bottleneck towards developing ML algorithms for providing quality care at a reduced cost.

\textbf{Background:} One way to circumvent this bottleneck is by simulating humans in appropriate virtual environments. The humans would generate physiological signals while the environment would be equipped with ambient sensors as in a smart home or care facility. This will allow generation of potentially infinite amounts of shareable data. Also, simulated humans will allow evaluation of artificial intelligence (AI) agents\footnote{An  agent is anything that can be viewed as perceiving its environment through sensors and acting upon that environment through actuators \cite{RussellNorvig2020}. Such agents, implemented in software, have been reported in our prior work \cite{chandrasekaran2004architecture,BanerjeePhDthesis2007,chandrasekaran2009diagrammatic,BanerjeeChandra2010JAIR,BanerjeeChandra2010,chandrasekaran2011augmenting,BanerjeeDutta2013BigDataSELP,BanerjeeDutta2013ICDMselp,BanerjeeDutta2014,najnin2016emergence,najnin2017predictive,najnin2018pragmatically,kapourchali2018multiple,kapourchali2020epoc,kapourchali2020state,baruah2020multimodal,baruah2020perception} as well as in others'.}, implemented in software, in terms of fidelity to human physiological performance. Virtual humans are ubiquitous. There also exist virtual patient simulators. However, a framework to derive physical and physiological signals from a virtual human round-the-clock as a function of his day-to-day actions and emotions is currently missing.



We aim to simulate a virtual human with wearable sensors that will synchronously generate five signals: skeletal joint motions, electrocardiogram (ECG), blood pressure (BP), respiration, and skin conductance response (SCR), as a function of an open-ended set of actions and emotions (e.g., walking sadly). That is, $P(S_t)=f_{\theta} (P(E_t ),P(A_t ))$, where $E_t$ and $A_t$ are the sets of emotions and actions respectively and $S_t$ is the set of sensor signals (skeletal motion, physiological signals) at time $t$; $f$ is a mapping from the probability distributions of $\{E_t, A_t\}$ to that of $S_t$; and $\theta$ are the parameters of $f$ that can be manipulated to generate signals for different emotions and actions with inter- and intra-individual variability.

\textbf{Contributions:} In this paper, we develop for the first time: \textbf{(1)} Computational models for synthesizing the five signals as a function of an open-ended set of actions and emotions. Extensive evaluations, involving user studies (MTurk), benchmark datasets (DEAP and HCI-Tagging for emotions, AmI and TROIKA for actions) and findings in the literature \cite{kreibig2010autonomic}, show that the models can synthesize the signals with high fidelity to real-world data. No model for synthesizing any of these signals takes into account actions and emotions jointly. \textbf{(2)} A modular system that combines the above models to synchronously synthesize the physiological signals from emotionally-expressive skeletal motion. Currently, there is no system even remotely related to this.

\section{Models and Methods} 
\label{Sec:Models and Methods}

\begin{figure*}[t!]
  \centering
  \vspace{-1mm}
  \subfigure[Overall block diagram of our system\label{Fig:BlockDiagram_overall}]{\includegraphics[width=0.7\textwidth]{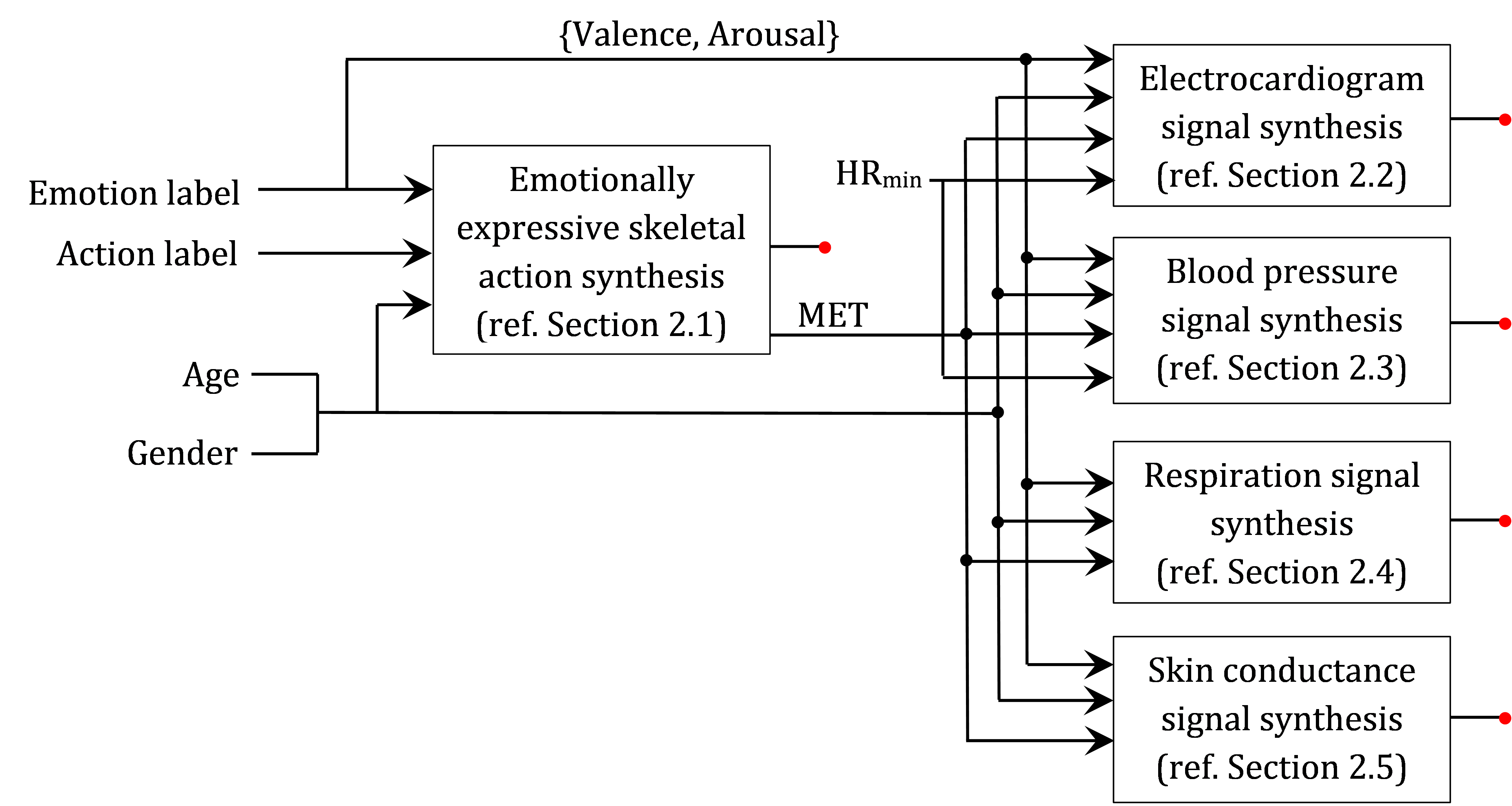}}\\
  \subfigure[Emotionally-expressive skeletal action synthesis (see Fig. \ref{Fig:Walk_Run_Neutral_Sad_Happy})\label{Fig:BlockDiagram_SkeletalAction}]{\includegraphics[width=0.58\textwidth]{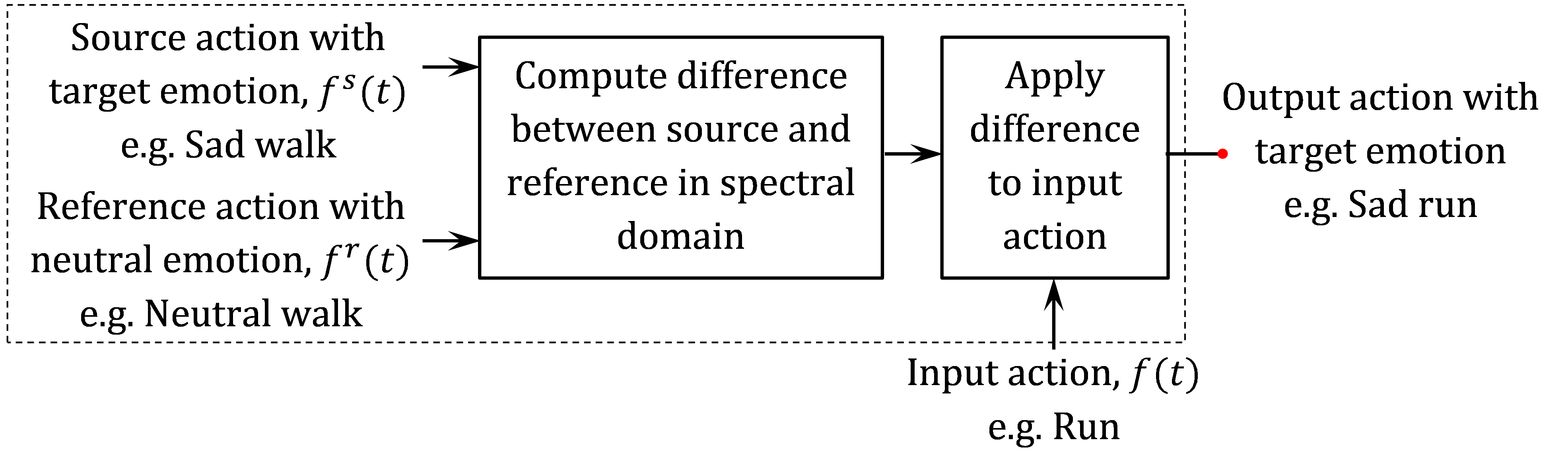}}\hfill
  \subfigure[ECG, BP signal synthesis (6 sec duration shown)\label{Fig:BlockDiagram_ECG+BP}]{\includegraphics[width=0.39\textwidth]{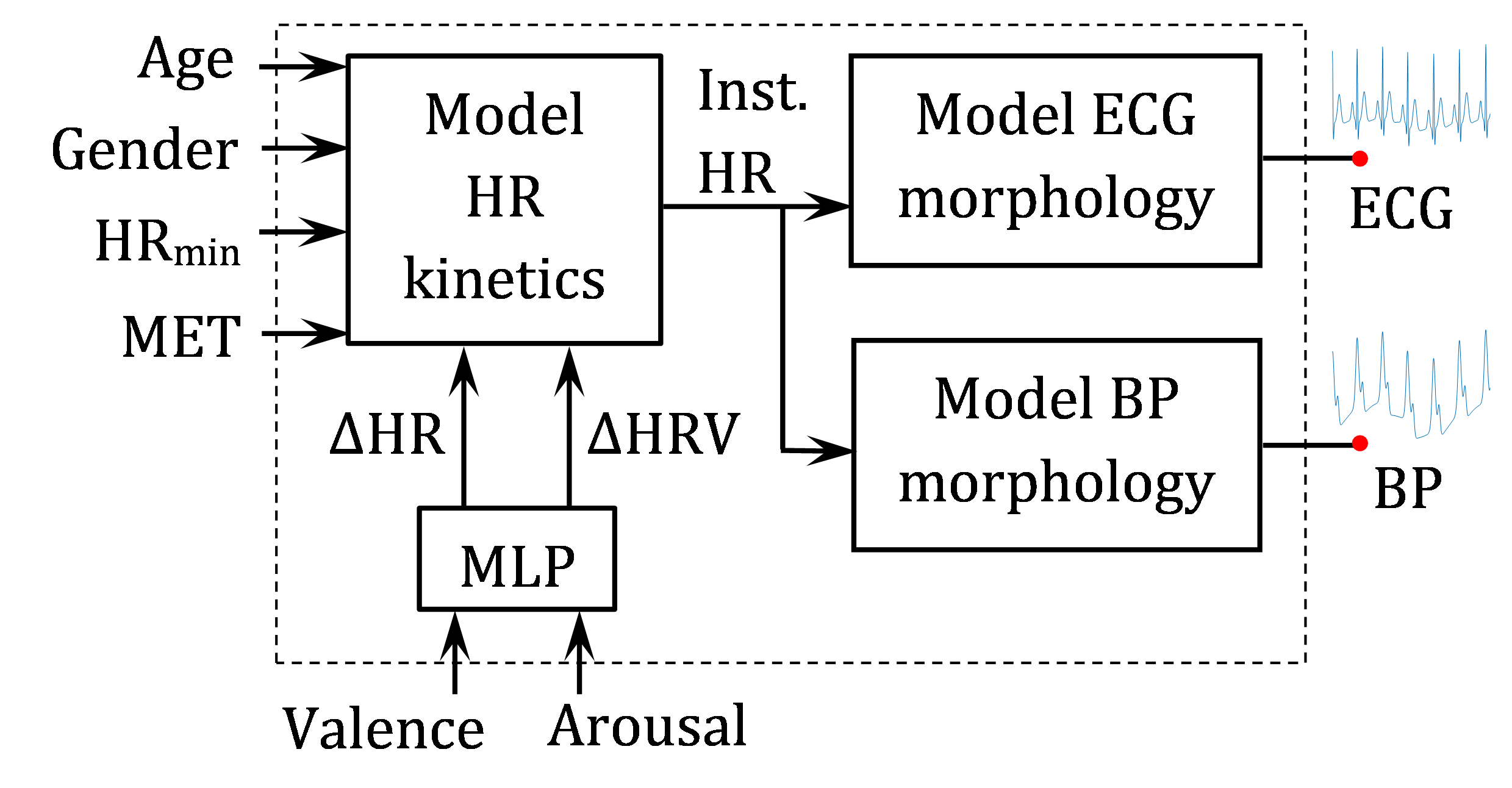}}\\
  \subfigure[SCR signal synthesis (60 sec duration shown)\label{Fig:BlockDiagram_SCR}]{\includegraphics[width=0.55\textwidth]{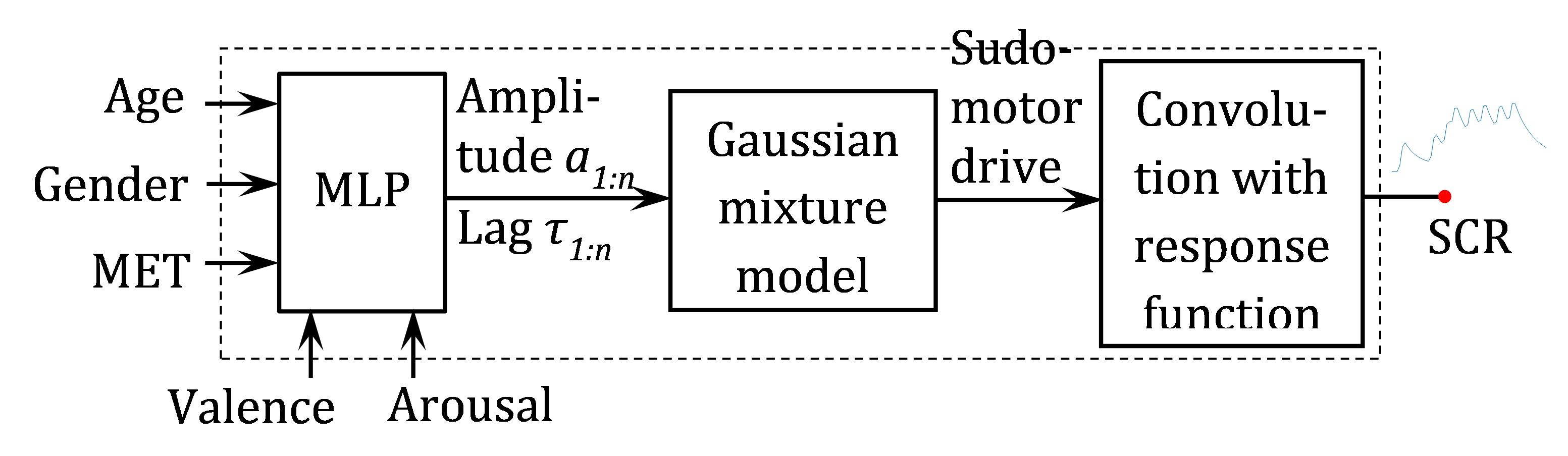}}\hfill
  \subfigure[Respiration signal synthesis (6 sec duration shown)\label{Fig:BlockDiagram_respiration}]{\includegraphics[width=0.41\textwidth]{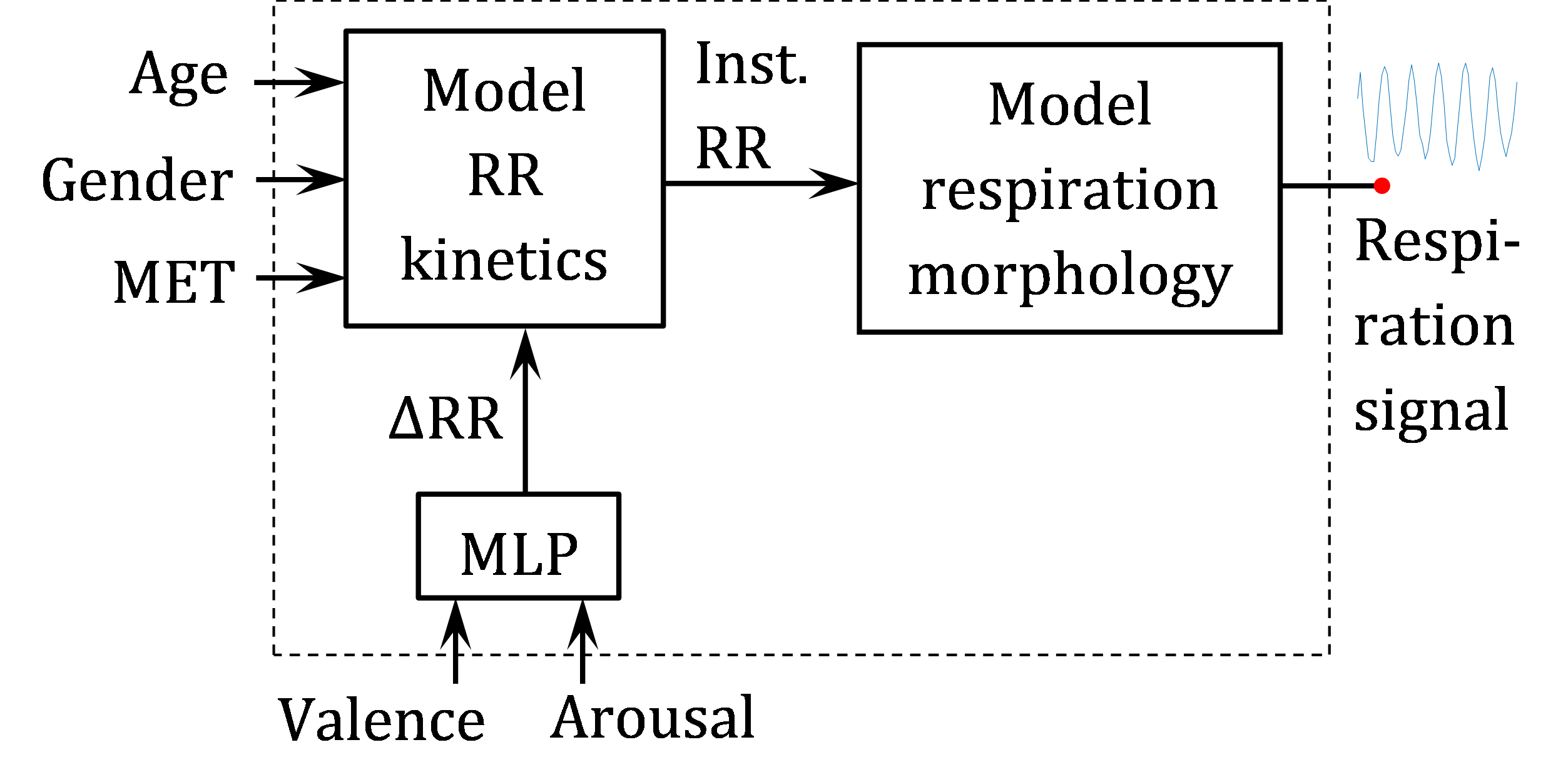}}
  \caption{Block diagrams of the proposed system and its components. ``Inst.'' refers to instantaneous. An input to all components is time which is not shown. The system can be tailored to each individual.}
  \vspace{-1mm}
\end{figure*}

Our system is shown in Fig. \ref{Fig:BlockDiagram_overall}. The synthesized skeletal motions generate a rate of energy expenditure in \emph{metabolic equivalent} (MET) \cite{ainsworth2000compendium_short} which is an input to the physiological signal models.

\subsection{Model for emotionally-expressive human skeletal motion synthesis.}
\label{Sec:Human skeletal motion generation}

The goal is to synthesize an extensive range of realistic skeletal actions, with individual variability, over a large range of emotions. Our idea is to consider an emotion as a style, and use a style transfer method to transfer an emotion from one action to another. This allows us to synthesize any action over any emotion as long as an instance of the action and the emotion are observed. See Fig. \ref{Fig:BlockDiagram_SkeletalAction} for an overview of our model. 

We use the spectral style transfer method in \cite{yumer2016spectral} which produces state-of-the-art results using a small and efficient model. Let $f^{s}[t]$ be a time domain signal of the source style, and a different action than the target motion $f[t]$. Let $f^{r}[t]$ be a reference signal that represents the same action as $f^{s}[t]$ and the same style as $f[t]$. The idea is to apply the difference between $f^{s}[t]$ and $f^{r}[t]$ to $f[t]$. Since the length, synchronization and spatial correspondences of the three signals may not match in the time domain, we formulate style transfer in the spectral domain by computing a new magnitude, $R'[\omega]$ for the entire signal, as: $R'[\omega] = R[\omega]+s[\omega](R^{s}[\omega] - R^{r}[\omega])$, $A'[\omega]=A[\omega]$, and $\texttt{Im}\{f'[t]\}=0$, over all $\omega$. Here, $s[\omega]=R[\omega] \big/ \displaystyle{\maxi_{\omega}}(R[\omega])$, $R(\omega) = |F(\omega)|$, $A(\omega) = \angle F(\omega)$, $F(\omega) = \mathcal{F}(f(t))$, and $\mathcal{F}$ denotes Fourier transform. The stylized magnitude $R'[\omega]$ and constant $A'[\omega]$ constitute the final signal, inverse Fourier transform of which is the stylized time domain output. See \cite{yumer2016spectral} for details.

\subsection{Electrocardiogram (ECG) signal synthesis model.}
\label{Sec:Electrocardiogram (ECG) model}

The ECG, BP and respiration signals vary in different ways with change in emotion and action. The cardiovascular and respiratory systems in the human body are intimately linked in their operations, hence they are often modeled together (e.g., \cite{clifford2004realistic}). Integrated models exploit the interdependencies between the two systems to synthesize more realistic signals. Closed-loop models receive internal feedback for stabilization of the systems, which also leads to more realistic signals. The proposed ECG, BP and respiration models are closed-loop and integrated.

The dynamical model for synthesizing ECG in \cite{mcsharry2003dynamical} was extended to synthesize BP and respiration signals \cite{clifford2004realistic}. It accurately reproduces many of the important clinical properties of these signals, such as QT dispersion, realistic beat-to-beat variability in timing and morphology as well as pulse transit time. It considers the balance between the effects of the sympathetic and parasympathetic systems. We extend these models \cite{mcsharry2003dynamical,clifford2004realistic} to incorporate the influence of actions and emotions.


The model \cite{mcsharry2003dynamical} generates a trajectory in 3D state space with coordinates $(x, y, z)$. Quasi-periodicity of the ECG is reflected by the movement of the trajectory around an attracting limit cycle of unit radius in the $(x, y)$-plane. Each revolution on this circle corresponds to one RRI\footnote{\{P, Q, R, S, T\} are the extrema points in one heartbeat in ECG signal. RR-interval (RRI) is the time between successive R peaks. HR stands for heart rate, HRV for heart rate variability.} or heart beat. Inter-beat variation in the ECG is reproduced using the motion of the trajectory in the $z$-direction. P, Q, R, S and T are described by events corresponding to negative and positive extrema in the $z$-direction. Motion is described by three ODEs: $\dot{x} = \alpha x-\omega y$, $\dot{y}=\alpha y+\omega x$, $\dot{z} = -\sum_{i\in\{\text{P,Q,R,S,T}\}} a_i\Delta\theta_i \exp(-\Delta\theta_i^2/2b_i^2)-(z-z_0)$, where $\alpha=1-\sqrt{x^2+y^2}$, $\Delta \theta_i=(\theta-\theta_i) \text{mod} 2\pi$, $\theta=\text{atan2}(y,x)$ and $\omega = 2\pi f$ is the angular velocity of the trajectory as it moves around the limit cycle and is related to the beat-to-beat HR. The baseline wander of the ECG is: $z_0(t) = A\text{sin}(2\pi f_2 t)$, where $A=0.15$ mV and $f_2$ is the respiratory frequency. The signals are scaled and shifted to the appropriate range. Table \ref{Table:parametersofECGandBP} lists the parameter values used in our ECG model (Fig. \ref{Fig:BlockDiagram_ECG+BP}).

HR tends to jump between quantized states, relating to different physical and mental activities \cite{clifford2004generating}. Since the model considers beat-to-beat HR dynamics, it is possible to add the effect of emotion and action to HR kinetics. Instead of using predefined Gaussians for RRI simulation as in \cite{mcsharry2003dynamical}, we use a mathematical model of HR kinetics to generate RRI time-series in response to actions and emotions.

The model of HR kinetics in response to movement \cite{zakynthinaki2015modelling} is expressed as a system of coupled ODEs: $\dot{\text{HR}} = -\textit{f}_{min}\textit{f}_{max}\textit{f}_{D}$, $\dot{v} = I(t)$, where HR is the subject's current heart rate, $t$ denotes time, $v$ is the movement intensity (velocity), $\lambda \in [0,1]$ is a global parameter that defines the overall cardiovascular condition of the subject and is a function of $\text{HR}_{min}$ and gender (ref. Eq. 6, 7 in \cite{zakynthinaki2015modelling}). $\lambda \approx 1$ corresponds to excellent cardiovascular condition. $\textit{f}_{min}$ controls the dynamics near the minimum HR ($\text{HR}_{min}$). $\text{HR}_{min}$ is a function of $\lambda$ and gender. $\textit{f}_{max}$ controls the dynamics near the maximum HR. $\text{HR}_{max}$ can be calculated as a function of age \cite{tanaka2001age}. $f_D$ is a function of initial HR, velocity, $\lambda$, blood lactate, and $t$.

Intensity of blood lactate is modeled as \cite{yang2017quantifying}: $I_l = I_p - arctan(I_p)$, where $I_p = v/v'_{max}$ is exercise intensity, $v'_{max}=40\sqrt{\lambda}$ is the maximum achievable velocity for a subject. The velocity of activities can be measured by 3D accelerometer along three dimensions as in \cite{yang2017quantifying}, or obtained as the mean velocity of points in a Kinect-type skeleton as in our work.

There have been efforts (e.g., \cite{yang2017quantifying}) to quantify mental arousal levels in daily living. HR in response to activity level is predicted using HR kinetic model. The additional heart rate (prediction error) is assumed to be a measure of mental arousal. However, only effect of arousal is taken into account. Furthermore, while a power function can be fitted to change in HR ($\Delta \text{HR}$) \cite{azarbarzin2014relationship}, it is not consistent with Kreibig table (Table 2 in \cite{kreibig2010autonomic}). So to incorporate emotions, we use a multilayer perceptron (MLP) to learn a mapping from \{valence, arousal\} to $\{\Delta \text{HR}, \Delta \text{HRV}\}$ from data of 62 subjects obtained from the DEAP \cite{koelstra2012deap} and HCI \cite{soleymani2012multimodal} datasets. The outputs of MLP are added to HR demand calculated for action using Karavonen's formula \cite{karvonen1988heart}. The final output is instantaneous HR time-series which can be converted to RRI time-series (RRI = 60/HR). The RRI is applied to the original model for generating ECG. Ten hidden units is found to be optimal.

\subsection{Blood pressure (BP) signal synthesis model.}
\label{Sec:Blood pressure (BP) model}

The dynamical model for generating ECG \cite{mcsharry2003dynamical} has been extended in \cite{clifford2004realistic} to include blood pressure in isolation from the ECG. BP waveform is generated with the same non-linear model but different parameters and angles (ref. Table \ref{Table:parametersofECGandBP}, Fig. \ref{Fig:BlockDiagram_ECG+BP}). Similar to ECG, changes in angular frequency in the $(x, y)$-plane (due to the RRI time-series) drive the beat-to-beat variations in timing and shape of the BP waveform. Since effect of emotion and action is incorporated in the RRI, there is no need to train a new model for BP. Many of the morphological changes in the ECG and BP waveforms are due to the geometrical structure of the model. Since the amplitude of BP waveforms includes information about systolic BP (SBP) and diastolic BP (DBP), the fact that the BP is linearly coupled to mean HR and hence inversely to mean of RRIs, is used in scaling. The scaling values are in the range $(110 - 30\times\overline{\text{RRI}}, 200 - 80\times\overline{\text{RRI}})$ where $\overline{\text{RRI}}$ is the mean of RRIs in a predefined duration of time.

\subsection{Respiration signal synthesis model.}
\label{Sec:Respiration model}

The block diagram of our model is shown in Fig. \ref{Fig:BlockDiagram_respiration}. The model for generating ECG \cite{mcsharry2003dynamical} has been extended in \cite{clifford2004realistic} to generate the respiration signal. The power spectrum of the RRIs is selected a priori and used to generate the respiratory signal. Effect of action is incorporated in RR using a function borrowed from \cite{tehrani1998optimal}. The criterion of minimum average respiratory work rate is used to determine the optimal values of respiratory frequency. RR is then calculated as:
\begin{equation*} \label{Equ:Resp:respRate}
    f = \bigg(-K'V_D + \sqrt{K'^2V_D^2+32K'K''V_D\dot V_A}\bigg)\bigg/16K''V_D
\end{equation*}
\noindent where $K'=5.1$ is the lung elastance and $K''=4.52$ is the air viscosity factor in the lung, both are constants, and dead space volume $V_D = 0.1698\dot V_A + 0.1587$. The input to the model is the alveolar ventilation, $\dot V_A$, that may be expressed as a function of different activities (exercise intensity or metabolic rate) as \cite{martin1987pulmonary}: $\dot V_A = 0.868\dot V_{CO_2}/\dot P_{ACO_2}$. $P_{ACO_2}$ is partial pressure of $\text{CO}_2$ which is constant during mild to moderate exercise \cite{martin1987pulmonary}, $V_{CO_2}$ is $\text{CO}_2$ output, and $V_{O_{2}max}$ is the maximum rate of oxygen consumption. $\dot V_{CO_2}$ is close to $\dot V_{O_2}$ \cite{noguchi1982breath}. Since we know how HR changes with exercise intensity (ref. model of HR kinetics in Sec. \ref{Sec:Electrocardiogram (ECG) model}), $\dot V_{O_2}$ is computed using HR as in \cite{swain1994target} and $V_{O_{2}max}$ is computed as in \cite{martin1987pulmonary}: $V_{O_2} = V_{O_{2}max}(\text{HR}/\text{HR}_{max} - 0.3718)/0.6463$, and
\begin{equation*}\label{Equ:Resp:VO2}
V_{O_{2}max} =
    \begin{cases}
      4.2 - 0.032 \times age \pm  0.4, & \text{for men}\\
      2.6 - 0.014 \times age \pm  0.4, & \text{for women}
    \end{cases}.
\end{equation*}

As with HR, emotion is incorporated by training a MLP which maps from \{valence, arousal\} to $\Delta$RR. The optimal number of hidden units is experimentally found to be six. Body movement and other factors introduce noise and trends in respiration signals collected using respiration belt, as in certain benchmark datasets (e.g., DEAP, HCI-Tagging). RR is calculated as the dominant frequency of a 6-sec sliding window over the de-trended and bandpass (0.1-0.9 Hz) filtered respiration signal.

\subsection{Skin conductance response (SCR) signal synthesis model.}
\label{Sec:Skin conductance response (SCR) model}

SCR generation models belong to three classes: curve fitting (e.g., \cite{lim1997decomposing}), peripheral (e.g., \cite{alexander2005separating,bach2010modelling}), and neural (e.g., \cite{bach2011dynamic,bach2010dynamic}). We consider a peripheral model in which the SCR is generated by convolving a response function, $h(t)$, and the sudomotor drive, $u(t,\Theta)$: $SCR(t) = h(t) \otimes u(t,\Theta)$. The skin conductance response function is approximated by a Gaussian smoothed biexponential function \cite{bach2010modelling}: $h(t) = N(t) \otimes E(t)$, where $N(t) = \frac{1}{\sqrt{2\pi}\sigma_{2}}\exp(-(t-t_0)^{2}/2\sigma_{2}^2)$, $E(t)=E_{1}(t)+E_{2}(t)$, $E_{i}(t)=\exp(-\lambda_{i}t)$, $i=1, 2$. The parameters $t_0=3.0745$ seconds for peak latency, $\sigma_{2}=0.7013$ for defining rise time, and $\lambda_{1}=0.3176$, $\lambda_{2}=0.0708$ for defining two decay components, as in \cite{bach2010modelling}.

The sudomotor drive, input to the skin conductance function, is modeled as mixture of Gaussians \cite{bach2011dynamic}: $u(t,\Theta) = \sum_{i=1}^n a_{i}\exp(-(t-\tau_{i})^{2}/2\sigma_{1}^{2})$, $\Theta = \{\tau_{i},a_{i}:i = 1, 2, \ldots, n\}$, where $n$ is the number of sudomotor nerve activity (SNA) bursts per minute, $\sigma_1$ is the standard deviation of the Gaussian, $a_{i}$ is the amplitude, and $\tau_i$ is the time of maximum firing of the SNA burst, $i$. While the duration and shape of SNA firing bursts are not well-defined, $n$ cannot exceed 30 bursts/min \cite{bach2011dynamic}. We assume $n=10$ bursts/min and $\sigma_1 = 0.3$, as in \cite{bach2011dynamic}.

The parameters $\Theta$ are generated as a function of emotion (valence, arousal), action (MET), age, gender and time using an MLP. The MLP weights are learned by minimizing $\sum_{m=1}^{M} \|SCR_{m}-SCR'_m\|^2$, where $SCR_{m}$ and $SCR'_{m}$ are the true and predicted SCR signals for the $m^{th}$ window. A window is of 60 seconds duration and slides by each second to generate a total of $M$ windows. We use $tanh$ activation function in the hidden layer (128 units) and $sigmoid$ in the output layer. The model is trained end-to-end using backpropagation and Adam optimizer with default hyperparameters $\beta_1 = 0.9$, $\beta_2 = 0.999$; learning rate of 0.001 and minibatch size of 100 for emotions, while 0.0001 and 500 for actions. See Fig. \ref{Fig:BlockDiagram_SCR} for a block diagram of our model.

\section{Empirical Evaluation} 
\label{Sec:Experimental Results}

Currently, there is no benchmark dataset containing physiological signals as a joint function of actions and emotions. Hence, our physiological signal synthesis models are evaluated for emotions assuming sitting action and for actions assuming neutral emotion. In all models, the hyperparameters are estimated from the training set via cross-validation. A simple model (MLP with one hidden layer) with regularization and dropout is used to prevent overfitting. Unless otherwise stated, in all experiments, randomly-chosen 60\% of data for each emotion or action in the dataset is used for training and the rest for testing.

\subsection{Evaluation of emotionally-expressive skeletal actions.}
\label{Sec:Evaluation of Emotionally-expressive Skeletal Actions}

\begin{figure*}[tb]
  \centering
  \vspace{-2mm}
  \hspace{-1cm}
  \subfigure{\includegraphics[width=0.71\textwidth]{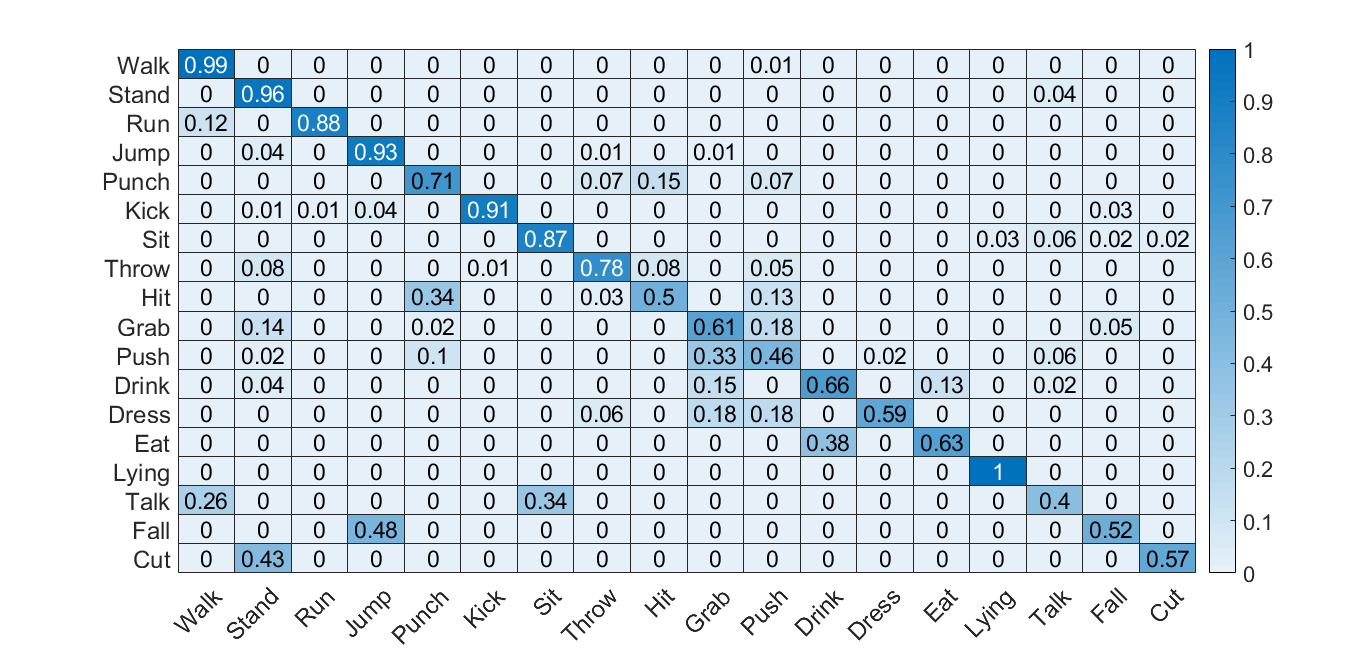}}\label{Tab:confusionActionRecognitionAccuracytable}\hfill
  \subfigure{\includegraphics[width=0.34\textwidth]{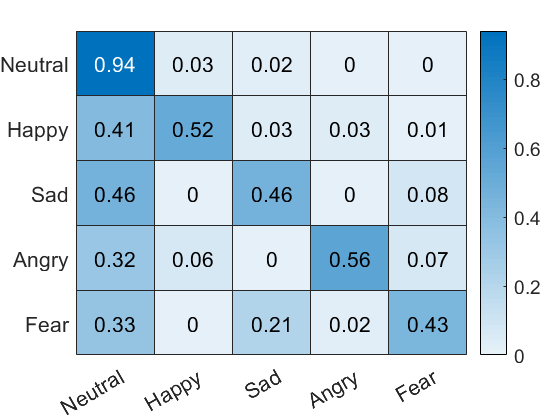}}\label{Tab:ConfusionEmotionRecognitionAccuracytable}\hspace{-1cm}\\
  \vspace{-2mm}
  \hspace{-1cm}
  \subfigure{\includegraphics[width=0.71\textwidth]{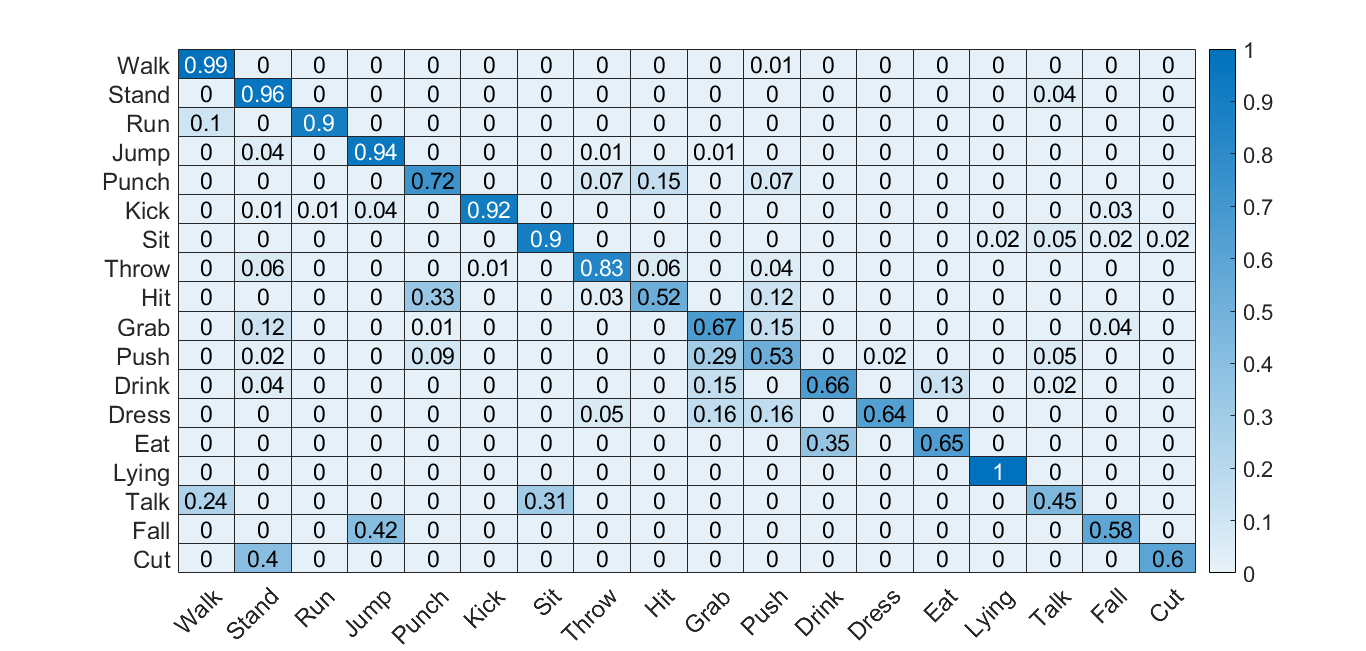}}\label{Tab:confusionActionRecognitionAccuracytable_benchmark}\hfill
  \subfigure{\includegraphics[width=0.34\textwidth]{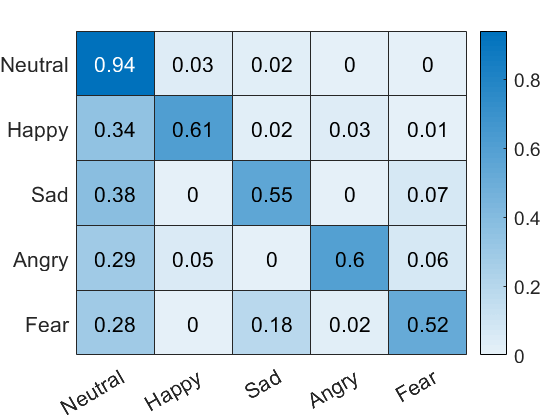}}\label{Tab:ConfusionEmotionRecognitionAccuracytable_benchmark}\hspace{-1cm}\\
  \caption{Confusion matrix for 18 skeletal actions (left) and five emotions (right) as obtained from a recognition task carried out by MTurk users on synthesized (top row) and benchmark (bottom row) data. The recognition rate is stated in each cell. The true and predicted classes are along the rows and columns respectively.}
  \label{Fig:SkeletalActionsEmotionsConfusionMatrices}
\end{figure*}

\textbf{Data:} We experimented with two benchmark datasets: CMU motion capture database \cite{cmu} in BVH format and Emotional Body Motion database \cite{ebmd}. The data has 38 skeleton joints with three degrees of freedom for each joint and joint movements are recorded at 120 frames per second. We used 18 actions and five emotions, which are listed in Fig. \ref{Fig:SkeletalActionsEmotionsConfusionMatrices}. Our model was tested on CMU mocap database.

\textbf{Experimental setup:} We conducted a user study on Amazon Mechanical Turk (MTurk) to assess how well our model generates emotionally expressive skeletal actions. Our user group consisted of MTurk Master workers; they have achieved distinction by consistently demonstrating a high degree of success in performing a large number and wide range of Human Intelligence Tasks. We administered two questions types to simultaneously assess action and emotion in videos of skeletal motions generated by our model. In one question type, we asked the users to choose one action out of 18, and in the second, to choose one emotion out of five, both from the same video. We created five survey forms, each containing 30 videos by combining 18 actions and five emotions, and a total of 60 questions. We rejected a response if it was recorded by a user before the video ended. If such an event occurred twice for an user, we rejected all his responses. After all rejections, we collected responses from 50 users, ten for each survey form. We conducted the same user study twice: once by presenting videos of emotionally expressive skeletal actions from the benchmark dataset, and another by presenting videos of the same but generated by our model. 

\textbf{Evaluation results:} The recognition rate from our user study for the videos generated by our model for the 18 actions and five emotions are shown in the confusion matrices in Fig. \ref{Fig:SkeletalActionsEmotionsConfusionMatrices}. The recognition rates for each action and emotion from our two user studies are very close. As expected, actions with similar skeletal motions such as hitting vs. throwing vs. pushing, eating vs. drinking, and falling vs. jumping, are confused by users. To the best of our knowledge, no user study has simultaneously evaluated the recognition of actions and emotions from the movements of human figures such as skeletons.

\begin{figure*}[tb]
  \centering
  \vspace{-2mm}  
  \subfigure[Neutral walk.\label{Fig:Neutral walk}]{\includegraphics[width=0.4\textwidth]{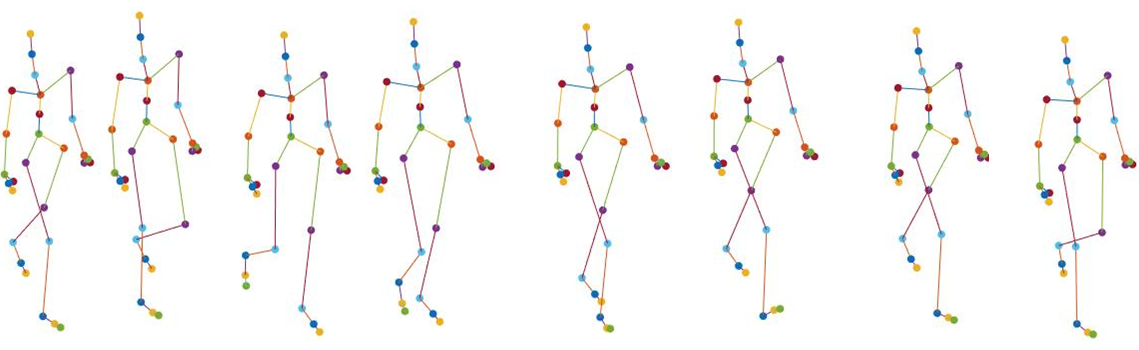}}\hfill
  \subfigure[Neutral run.\label{Fig:Neutral run}]{\includegraphics[width=0.52\textwidth]{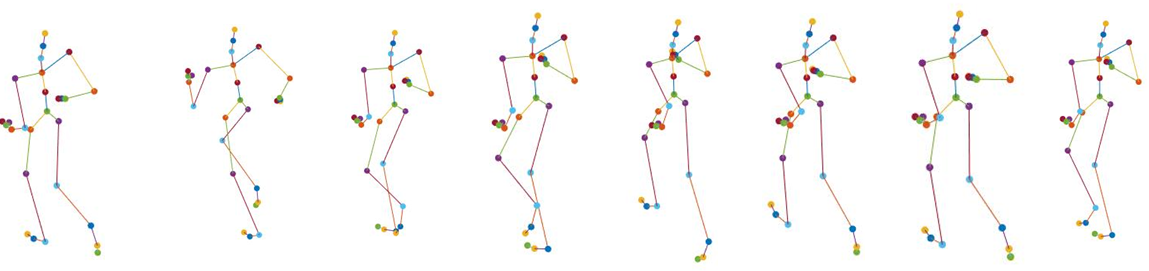}}\\
  \vspace{-1mm}  
  \subfigure[Sad walk.\label{Fig:Sad walk}]{\includegraphics[width=0.4\textwidth]{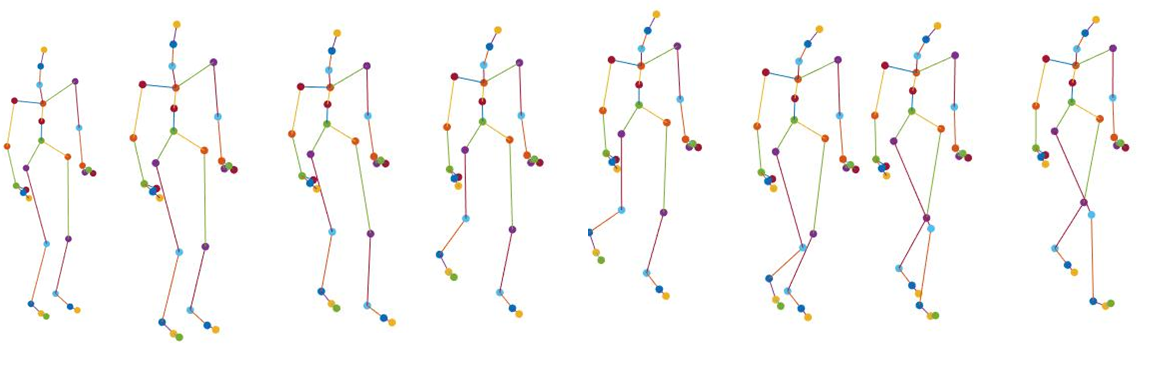}}\hfill
  \subfigure[Happy run.\label{Fig:Happy run}]{\includegraphics[width=0.52\textwidth]{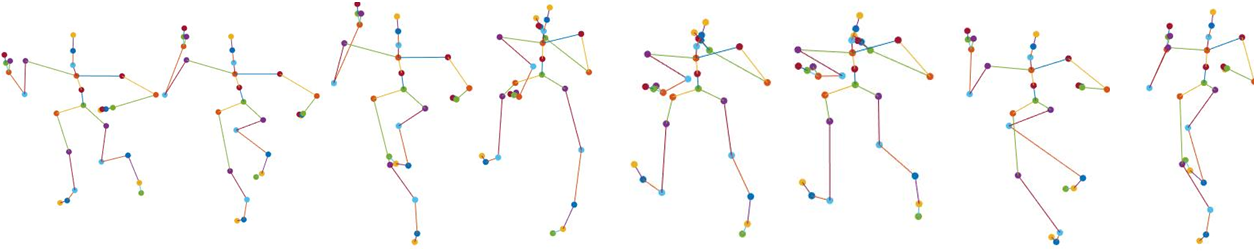}}\\
  \caption{Visual comparison of key frames of walking (left) and running (right) for different emotions. The top row (a or b) and an emotion label (e.g., sad, happy) are input to our model while the corresponding output is the bottom row (c or d).} 
  \label{Fig:Walk_Run_Neutral_Sad_Happy}
\end{figure*}

In Fig. \ref{Fig:Walk_Run_Neutral_Sad_Happy}, skeletal motions for two sample actions, walk and run, are shown for neutral and happy/sad emotions. Consistent with the literature \cite{melzer2019we}, for sad emotion, skeleton head joints are having slow, long movements over time with low amplitude elbow motion and a hiding posture. For happy emotion, the arms are stretched out to the front.

\subsection{Evaluation of physiological signals on emotions.}
\label{Sec:Evaluation of physiological signals on emotions}


\textbf{Datasets:} The physiological signals, generated by our models, are evaluated as a function of emotion using the Kreibig table and two benchmark datasets.

Kreibig \cite{kreibig2010autonomic} reviewed 134 publications reporting experimental investigations of emotional effects on peripheral physiological responding. Table 2 in \cite{kreibig2010autonomic} summarizes the findings. It shows the response direction reported by majority of studies with at least three studies indicating the same response direction. It does not include the response magnitude since magnitude quantification depends on the type of baseline or comparison condition which varies greatly across studies.

DEAP \cite{koelstra2012deap} is a dataset for emotion analysis using physiological signals. It contains ECG, BP, respiration and SCR, recorded from 32 participants as each watched 40 excerpts of music videos of one minute duration. Participants rated each video in terms of the levels of arousal, valence, like/dislike, dominance and familiarity.

HCI Tagging \cite{soleymani2012multimodal} is a database of user responses in terms of peripheral physiological signals and self-annotations to multimedia content. Twenty seven participants (11 male, 16 female, aged 19-40 years) were shown 20 fragments of movies, while recording their physiological signals such as ECG, respiration amplitude, SCR, and others. For each video shown, the participants self-reported their emotive state on a discrete scale of 1-9 for valence and arousal along with an emotional keyword.


\textbf{Evaluation results:} The training set is chosen such that the signal direction for an emotion in that set matches with the direction in the Kreibig table. Our models synthesized the four signals for the emotions in Kreibig table and compared the directions of change from neutral emotion using the following features: HR and HRV for ECG; SBP, DBP and LVET (left ventricular ejection time) for BP; RR for respiration; and SCR. See Table \ref{Table:ECG_BP_RRvsKreibig}. The valence and arousal values of these emotions are obtained from \cite{scherer2005emotions}. A threshold of 10\% of maximum and minimum of changes is used to obtain the directions. Four mismatches occurred out of 64 entries in Table \ref{Table:ECG_BP_RRvsKreibig}, leading to an error rate of 6.25\%.

To evaluate the performance of our models in predicting the signal direction (increase, decrease, no change) due to change in emotions using benchmark datasets, direction of the synthesized signal is compared with that of the true signal. As above, a 10\% threshold is used to obtain the directions. See Table \ref{tab:emotionData}. Direction of signals due to emotions in benchmark datasets often do not match with that in the Kreibig table due to significant variability between individuals. Thus, if a model is trained on the Kreibig table, its performance may be lower on the benchmark datasets, and vice versa.

\begin{table*}[t]
\caption {Comparison of feature directions extracted from our synthesized ECG (HR, HRV), BP (SBP, DBP, LVET), RR and SCR signals with those in Kreibig table. Arrows indicate increase ($\uparrow$), decrease ($\downarrow$), no change from baseline (-), or both increase and decrease ($\uparrow\downarrow$) between studies. Arrows in parentheses indicate tentative response direction, based on fewer than three studies. Mismatched directions are marked red.} 
\label{Table:ECG_BP_RRvsKreibig}
\begin{center}
 \begin{tabular}{|c|c|c|c|c|cc|c|c|cc|cc|c|c|c|c|c|}
 \hline
 \rotatebox{90}{Signal} & \rotatebox{90}{Source of} \rotatebox{90}{direction} & \rotatebox{90}{Anger} & \rotatebox{90}{Anxiety} & \rotatebox{90}{Disgust} & \multicolumn{2}{|c|}{\rotatebox{90}{Embarrass-} \rotatebox{90}{ment}} & \rotatebox{90}{Fear} & \rotatebox{90}{Sadness} & \multicolumn{2}{|c|}{\rotatebox{90}{Amuse-} \rotatebox{90}{ment}} & \multicolumn{2}{|c|}{\rotatebox{90}{Content-} \rotatebox{90}{ment}} & \rotatebox{90}{Happiness} & \rotatebox{90}{Joy} & \rotatebox{90}{Suspense} & \rotatebox{90}{Pleasure} & \rotatebox{90}{Relief}\\
  \hline
  \hline
  \multirow{2}{*}{HR} & Kreibig table & $\uparrow$ & $\uparrow$ & $\uparrow$- & \multicolumn{2}{|c|}{$\uparrow$} & $\uparrow$ & $\downarrow$ & \multicolumn{2}{|c|}{$\uparrow\downarrow$} & \multicolumn{2}{|c|}{$\downarrow$} & $\uparrow$ &  $\uparrow$ &  $(\downarrow$) & & \\ 
  \cline{2-18}
   & Synthesized & $\uparrow$ & $\uparrow$ & $\uparrow$- & \multicolumn{2}{|c|}{$\uparrow$} & $\uparrow$ & $\downarrow$ & \multicolumn{2}{|c|}{$\uparrow$} & \multicolumn{2}{|c|}{$\downarrow$} & $\uparrow$ & $\uparrow$ & $\downarrow$ & & \\
  \hline
  \multirow{2}{*}{HRV} & Kreibig table & $\downarrow$ & $\color{red}\downarrow$ & $\uparrow $ & \multicolumn{2}{|c|}{$\color{red}(\downarrow)$} & $\downarrow$ & $\downarrow$ & \multicolumn{2}{|c|}{$\uparrow$} & \multicolumn{2}{|c|}{$\uparrow\downarrow$} & $\downarrow$ & $\color{red}(\uparrow)$ & & & \\
  \cline{2-18}
   & Synthesized & $\downarrow$ & $\color{red}\uparrow$ & $\uparrow $ & \multicolumn{2}{|c|}{$\color{red}\uparrow$} & $\downarrow$ & $\downarrow$ & \multicolumn{2}{|c|}{$\uparrow$} & \multicolumn{2}{|c|}{$\downarrow$} & $\downarrow$ & $\color{red}\downarrow$ & $\downarrow$ & & \\
  \hline
  \hline
  \multirow{2}{*}{SBP} & Kreibig table & $\uparrow$ & $\uparrow$ & $\uparrow $ & \multicolumn{2}{|c|}{($\uparrow$)} & $\uparrow$ & & \multicolumn{2}{|c|}{$\uparrow $-} & \multicolumn{2}{|c|}{($\downarrow$)} & $\uparrow$&  $\uparrow$ & & & \\
  \cline{2-18}
   & Synthesized & $\uparrow$ & $\uparrow$ & $\uparrow $ & \multicolumn{2}{|c|}{$\uparrow$} & $\uparrow$ & $\downarrow$ & \multicolumn{2}{|c|}{$\uparrow$} & \multicolumn{2}{|c|}{$\downarrow$} & $\uparrow$ & $\uparrow$ & $\downarrow$ & & \\
  \hline
  \multirow{2}{*}{DBP} & Kreibig table & $\uparrow$ & $\uparrow$ & $\uparrow $ & \multicolumn{2}{|c|}{($\uparrow$)} & $\uparrow$ & & \multicolumn{2}{|c|}{$\uparrow $-} & \multicolumn{2}{|c|}{($\downarrow$)} & $\uparrow$&  - & & & \\
  \cline{2-18}
   & Synthesized & $\uparrow$ & $\uparrow$ & $\uparrow $ & \multicolumn{2}{|c|}{$\uparrow$} & $\uparrow$ & $\downarrow$ & \multicolumn{2}{|c|}{$\uparrow$} & \multicolumn{2}{|c|}{$\downarrow$} & $\uparrow$ &  $\uparrow$ &  $\downarrow$ & & \\
  \hline
  \multirow{2}{*}{LVET} & Kreibig table & $\downarrow$ &  & ($\downarrow $) & \multicolumn{2}{|c|}{} & $\downarrow$ &   & \multicolumn{2}{|c|}{} & \multicolumn{2}{|c|}{($\uparrow$)} & \color{red}(-) & ($\downarrow$) & & & \\
  \cline{2-18}
   & Synthesized & $\downarrow$ & $\downarrow$ & $\downarrow $ & \multicolumn{2}{|c|}{$\downarrow$} & $\downarrow$ & $\uparrow$ & \multicolumn{2}{|c|}{$\downarrow$} & \multicolumn{2}{|c|}{$\uparrow$} & $\color{red}\downarrow$ & $\downarrow$ & $\uparrow$ & & \\
  \hline
  \hline
  \multirow{2}{*}{RR} & Kreibig table & $\uparrow$ & $\uparrow$ & $\uparrow$ & \multicolumn{2}{|c|}{} & $\uparrow$ & $\uparrow$ & \multicolumn{2}{|c|}{$\uparrow$} & \multicolumn{2}{|c|}{$\uparrow\downarrow$} & $\uparrow$& $\uparrow\downarrow$ &($\uparrow$) & & \\
  \cline{2-18}
   & Synthesized & $\uparrow$ & $\uparrow$ & $\uparrow$ & \multicolumn{2}{|c|}{$\uparrow$} & $\uparrow$ & $\uparrow$ & \multicolumn{2}{|c|}{$\uparrow$} & \multicolumn{2}{|c|}{$\downarrow$} & $\uparrow$ & $\uparrow$ & $\uparrow$ & & \\
  \hline
  \hline
  \multirow{2}{*}{SCR} & Kreibig table & $\uparrow$ & $\uparrow$  & $\uparrow$  & \multicolumn{2}{|c|}{} & $\uparrow$ & $\downarrow$ & \multicolumn{2}{|c|}{$\uparrow$} & \multicolumn{2}{|c|}{(...)} & & & & $\uparrow$ & $\downarrow$\\
  \cline{2-18}
   & Synthesized & $\uparrow$ & $\uparrow$  & $\uparrow$  & \multicolumn{2}{|c|}{} & $\uparrow$  & $\downarrow$  & \multicolumn{2}{|c|}{$\uparrow$}  & \multicolumn{2}{|c|}{(...)} & & & & $\uparrow$ & $\downarrow$\\
  \hline
 \end{tabular}
\end{center}
\end{table*}

\begin{table*}[tb]
\parbox{.45\linewidth}{
 \captionof{table}{Parameter values used in our ECG (upper block) and BP (lower block) synthesis models. Time is in secs, $\theta_i$ in radians. \label{Table:parametersofECGandBP}}
\centering
 \begin{tabular}{|c|c|c|c|c|c|}
 \hline
  & P & Q & R & S & T\\
 \hline
 \hline
 Time & $-0.2$ & $-0.05$ & 0 & 0.05 & 0.3\\
 $\theta_i$ & $-\pi/3$ & $-\pi/12$ & 0 & $\pi/12$ & $\pi/3$\\
 $a_i$ & 1.2 & $-5$ & 30 & $-7.5$ & 0.75\\
 $b_i$ & 0.25 & 0.1 & 0.1 & 0.1 & 0.4\\
 \hline
 \hline
 Time & 0.21 & 0.01 & 0 & 0.03 & 0.22\\
 $\theta_i$ & $-5\pi/12$ & $-\pi/36$ & 0 & $\pi/18$ & $4\pi/9$\\
 $a_i$ & 0 & 0 & 0.45 & 0.25 & 0.45\\
 $b_i$ & 0.25 & 0.1 & 0.3 & 0.5 & 0.3\\
 \hline
 \end{tabular}
}
\hfill
\parbox{.45\linewidth}{
 \captionof{table}{Accuracy (\%) of our models for predicting the direction of signal movement as a function of emotions in DEAP and HCI data.\label{tab:emotionData}}
\centering
 \begin{tabular}{|c|c|c|c|c|}
 \hline
  &DEAP&DEAP&HCI&HCI\\
  &Female&Male&Female&Male\\
  \hline
  \hline
  HR &49.66&43.58&56.17&42.69\\
  \hline
  HRV &48&37.5&55.29&35.38\\
  \hline
  \hline
  SBP &47.35&48.66&-&-\\
  \hline
  DBP &42.05&47&-&-\\
  \hline
  LVET &52&51.66&-&-\\
  \hline
  \hline
  RR &52.17&59.56&60.59&53.08\\
  \hline
  \hline
  SCR &74.51&89.49&76.04&79.44\\
  \hline
 \end{tabular}
}
\end{table*}

\subsection{Evaluation of physiological signals on actions.}
\label{Sec:Evaluation of physiological signals on actions}

\textbf{Datasets:} The physiological signals, generated by our models, are evaluated as a function of action using two benchmark datasets.

AmI \cite{AmI} is a dataset for action analysis using physiological signals. AmI contains respiration and SCR, recorded from 10 participants. Their age and gender are not specified. Our models are simulated with their average age (27.2 years) and male gender. Participants did activities such as lying, walking, sitting, standing, kneeling, all fours (crawling), transition, leaning, cycling and running.

TROIKA \cite{zhang2015troika} is a dataset for heart rate monitoring when motion artifact is strong. It includes ECG, recorded from chest using wet ECG sensors from 12 male participants as each walked or ran on a treadmill at six different speeds, each for 0.5-1 minute. The participants are aged 18-35 years. Each individual's age is unknown, so we used their mean age.


\textbf{Evaluation results:} To evaluate our ECG model using TROIKA, a signal corresponding to each subject is generated (different initial heart rates and $\lambda$ lead to different signals for each individual). The Pearson correlation between synthesized and actual HR for each subject is significant $(p<0.01)$ with mean 0.72 (std. dev. 0.08). Our respiration and SCR models are evaluated using AmI following the same procedure as above. The Pearson correlation between synthesized and actual RR for each of the 10 subjects is significant with mean 0.82 (std. dev. 0.18). The same for SCR is also significant with mean 0.57 (std. dev. 0.23). We did not find any action dataset containing BP signals.

 \begin{table}[htbp]
  \caption{Multimodal classification (F1-score, precision, recall, \% accuracy) of true and synthesized signals. AmI baseline is from a support vector machine classifier.}\label{Table:Multimodal classification}
  \centering
  \begin{tabular}{|c|c|c|c|c|c|c|}
  \hline
  \multicolumn{2}{|c|}{\multirow{2}{*}{Dataset}} & \multirow{2}{*}{F1} & \multirow{2}{*}{Prec.} & \multirow{2}{*}{Rec.} & \multirow{2}{*}{Acc.} & Base\\
  \multicolumn{2}{|c|}{} &       &       &       &       &  Acc.  \\ \hline
  DEAP     & true  & 0.6   & 0.61  & 0.58  & 65.2  & 57 \\ \cline{2-6}
  arousal  & syn.  & 0.48  & 0.47  & 0.49  & 57.3  & \cite{koelstra2012deap} \\ \hline
  DEAP     & true  & 0.63  & 0.6   & 0.58  & 63.6  & 62.7 \\ \cline{2-6}
  valence  & syn.  & 0.49  & 0.48  & 0.5   & 55.3  & \cite{koelstra2012deap} \\ \hline
  HCI      & true  & 0.67  & 0.67  & 0.68  & 68.6  & 54.7 \\ \cline{2-6}
  arousal  & syn.  & 0.46  & 0.46  & 0.46  & 47.7  & \cite{wiem2017emotion} \\ \hline
  HCI      & true  & 0.66  & 0.65  & 0.65  & 65.6  & 56.8 \\ \cline{2-6}
  valence  & syn.  & 0.47  & 0.47  & 0.47  & 47.8  & \cite{wiem2017emotion} \\ \hline
  \multirow{2}{*}{AmI}& true  & 0.68  & 0.68  & 0.7   & 75.1  & 64.4 \\ \cline{2-6}
           & syn.  & 0.78  & 0.78  & 0.8   & 80.7  & (SVM)\\ \hline
 \end{tabular}
 \end{table}

\subsection{Downstream task.} 
We provide the baseline for multimodal action and emotion classification using the synthesized signals. We classify the true and synthesized signals separately using MLP and early fusion. For DEAP, valence and arousal are classified separately into two classes (high, low) using six features extracted from SCR, respiration and BP signals \cite{tripathi2017using}: mean and standard deviation of raw signals, mean of absolute of first and second differences of raw signals and the same of normalized signals, from a 6-second window with 50\% overlap. Each window constitutes a datapoint. For HCI, three-class (high, medium, low) classification is done using 13 features extracted from ECG, SCR and respiration signals \cite{tripathi2017using}: the six features as in DEAP, mean, standard deviation, median, minimum, maximum, range, variance, skewness and kurtosis from a 6-second window with 50\% overlap. We classify the AmI and synthesized signals into 10 actions using 13 features (same as HCI) extracted from SCR, HR and RR signals, from a 6-minute window with 50\% overlap. Results (ref. Table \ref{Table:Multimodal classification}) show that it is slightly easier to classify the true signals than the synthesized ones for emotions, and vice versa for actions.

\section{Conclusions} 
\label{Sec:Conclusions}

We developed a novel system for synchronously synthesizing skeletal motion and physiological signals (ECG, BP, respiration, SCR) as a joint function of a set of actions and emotions. This set is open-ended as the constituent models operate on \{valence, arousal\} values for emotion and MET value (derived from skeletal joint movements) for action. Extensive evaluation involving user studies, benchmark datasets and comparison to findings reported in the literature show that the proposed system can synthesize the signals with high levels of accuracy, for both actions and emotions. 

\textbf{Significance:} A modular system for synthesizing realistic motion and physiological signals from an individual's actions and emotions will: \textbf{(1)} overcome the five challenges stated at the beginning of this paper, thereby facilitating ML research for round-the-clock monitoring at a reduced cost; \textbf{(2)} allow reusability of code and data; \textbf{(3)} be useful for training ML students, data scientists, modelers, and healthcare professionals by facilitating simulation of critical situations, education, research, and in silico clinical trials; and \textbf{(4)} allow evaluation of AI agents in terms of fidelity to human physiological performance.

\textbf{Impact:} Using this system, it will be possible for the first time to generate infinite amounts of shareable data as a function of any action and emotion in any environment, without any risk of privacy invasion. This is the first dataset for joint action and emotion recognition from motion and physiological signals. Our system will open up a new line of research in signal synthesis as a joint function of actions and emotions as there is currently no system even remotely related to ours.

\bibliographystyle{unsrt} 
\bibliography{/Bonny/Research/LatexDocs/mybibfile,ECG_BP_Respiration_short,Reference_SkeletalMotion,SkinConductanceReferences_short,Reference_debu}

\end{document}